\documentclass[10pt,conference]{IEEEtran}\IEEEoverridecommandlockouts
\usepackage{epsfig,graphicx,subfigure,psfrag,amsmath,cases}
\usepackage{latexsym,amssymb,amsmath,epsfig,subfigure,algorithm,mathtools}
\usepackage{algorithmic}
\usepackage{color}
\usepackage{url}
\usepackage{scrtime}
\usepackage{array}
\author{\authorblockN{Derrick Wing Kwan Ng, Yan Sun,  and Robert Schober\thanks{Derrick Wing Kwan Ng and Robert Schober are also with the University of British Columbia, Canada. This work was supported by the AvH Professorship Program of the Alexander von Humboldt Foundation. }}

Institute for Digital Communications\\ Friedrich-Alexander-University Erlangen-N\"urnberg (FAU), Germany

}

\title{ Power Efficient and Secure  Full-Duplex Wireless Communication Systems
}

\newtheorem{Thm}{Theorem}

\newtheorem{T-Prob}{Transformed Problem}

\DeclareMathOperator{\Tr}{\mathrm{Tr}}
\DeclareMathOperator{\zero}{\mathbf{0}}
\DeclareMathOperator{\Rank}{\mathrm{Rank}}

\DeclareMathOperator{\mino}{minimize}

 \newcommand{\qed}{\hfill \ensuremath{\blacksquare}}
\newtheorem{Remark}{Remark}

\newcommand{\abs}[1]{\lvert#1\rvert}
\newcommand{\norm}[1]{\lVert#1\rVert}
\begin{document}
\IEEEspecialpapernotice{(Invited Paper)}
\maketitle

\begin{abstract}
In this paper, we study resource allocation  for a full-duplex (FD) radio base station serving  multiple half-duplex (HD) downlink and uplink users simultaneously. The considered resource allocation algorithm design is formulated as a non-convex optimization problem taking into account   minimum required receive signal-to-interference-plus-noise ratios (SINRs) for downlink and uplink communication and maximum tolerable SINRs at potential eavesdroppers. The proposed optimization framework enables secure downlink and uplink communication  via  artificial noise generation in the downlink for interfering the potential eavesdroppers.  We minimize the weighted sum of the total downlink and uplink transmit power by jointly optimizing the downlink beamformer, the artificial noise covariance matrix,  and the uplink transmit power.
 We adopt a semidefinite programming (SDP) relaxation  approach to obtain a tractable solution for the considered problem. The tightness of the SDP relaxation is revealed by examining a sufficient condition for the global optimality of the solution.  Simulation results demonstrate the excellent performance achieved by the proposed  scheme and the significant transmit power savings enabled  optimization of the artificial noise covariance matrix.
\end{abstract}
\renewcommand{\baselinestretch}{0.913}
\normalsize

\section{Introduction}
\label{sect1}
The development of wireless communication networks worldwide has triggered
an exponential growth in the number of wireless communication devices for applications such as e-health,  energy management, and safety management. It is expected that by $2020$,
the number of interconnected devices on the planet may reach up to $50$ billion. In response to the resulting tremendous energy and bandwidth consumption, recent efforts for next generation communication system development have aimed at
providing secure and high speed communication with guaranteed
quality of service (QoS). In particular,   multiple-input multiple-output (MIMO) full-duplex (FD) wireless communications has recently received significant attention from both academia and industry \cite{JR:Full_duplex_radio2}--\nocite{JR:Full_duplex_radio3,JR:Kwan_FD,JR:FD_DC_program2,JR:FD_antenna_selection}\cite{JR:FD_large_antennas}. In contrast to conventional half-duplex (HD) transmission, FD operation enables simultaneous downlink and uplink transmission  at the same frequency. In other words, it can potentially double the spectral efficiency of existing HD wireless communication systems.  In \cite{JR:Full_duplex_radio2}, the authors studied  the mitigation of self-interference in
FD MIMO relays. In \cite{JR:Full_duplex_radio3}, the outage probability of MIMO FD single-user relaying systems
was investigated. In \cite{JR:Kwan_FD}, the authors proposed a polynomial time computational complexity resource allocation algorithm  for data rate maximization in
 multicarrier multiuser MIMO FD relaying systems. In \cite{JR:FD_DC_program2}, a suboptimal beamformer design was studied to improve the spectral efficiency of 
FD radio base stations (BSs) enabling simultaneous uplink and downlink communication. Joint antenna selection and power allocation was investigated in \cite{JR:FD_antenna_selection} for distributed antenna systems for power efficiency FD communication. In \cite{JR:FD_large_antennas}, massive MIMO was exploited to  suppress the self-interference  in FD communication for enhancing the system throughput. In general, FD systems can serve more users simultaneously  compared to HD systems \cite{JR:Kwan_FD,JR:FD_DC_program2}. However, this may also increase
the susceptibility to eavesdropping as there are more opportunities for information leakage.

Security is a fundamental problem in wireless communication systems due to the
broadcast nature of the wireless medium. Traditionally, cryptographic encryption
technologies have been used to enable communication security in the application
layer. However, the commonly used encryption algorithms are based on
the assumption of limited computational capabilities at the eavesdroppers which
may not hold in the future due to the development of quantum computers.
 As an alternative, physical (PHY) layer security utilizes the physical
properties of wireless communication channels, such as  interference  and channel fading, to ensure perfectly secure communication \cite{JR:PHY_SEC_tutorial}-\nocite{Report:Wire_tap,JR:EE-sec,JR:Artifical_Noise1}\cite{JR:Kwan_physical_layer}, regardless of the
potentially unlimited computational capabilities of the potential eavesdroppers. In his pioneering work on PHY layer security \cite{Report:Wire_tap}, Wyner showed
   that a source and a destination can exchange perfectly secure
 messages at a strictly positive data rate  when the  source-user channel enjoys a better quality than the  source-eavesdropper channel. Hence, the spatial degrees of freedom offered by multiple antennas may be exploited to secure communication systems. In \cite{JR:EE-sec}, energy-efficient optimization for PHY layer security in
multi-antenna downlink networks was studied.   In \cite{JR:Artifical_Noise1} and
 \cite{JR:Kwan_physical_layer}, different artificial noise based
  power allocation algorithms were proposed for the maximization of the ergodic secrecy capacity and the outage secrecy capacity,  respectively. However, HD operation was assumed in these works and  the  results obtained in \cite{JR:PHY_SEC_tutorial}-\nocite{Report:Wire_tap,JR:EE-sec,JR:Artifical_Noise1}\cite{JR:Kwan_physical_layer} may not be applicable to the case of simultaneous downlink and uplink communication enabled by FD radio BSs.

Motivated by the aforementioned prior works, in this paper, we study the resource allocation algorithm design for multiuser FD wireless communication systems. We minimize the weighted sum of the downlink (DL) and uplink (UL) transmit powers  while   ensuring  the QoS of both UL and DL users for secure and reliable communication.  In particular, we propose a semidefinite programming (SDP) based resource allocation algorithm to obtain the optimal system performance.

\section{System Model}
In this section, we present the adopted  channel model for secure simultaneous DL and UL  communication.
\begin{figure}
\centering
\includegraphics[width=3.5in]{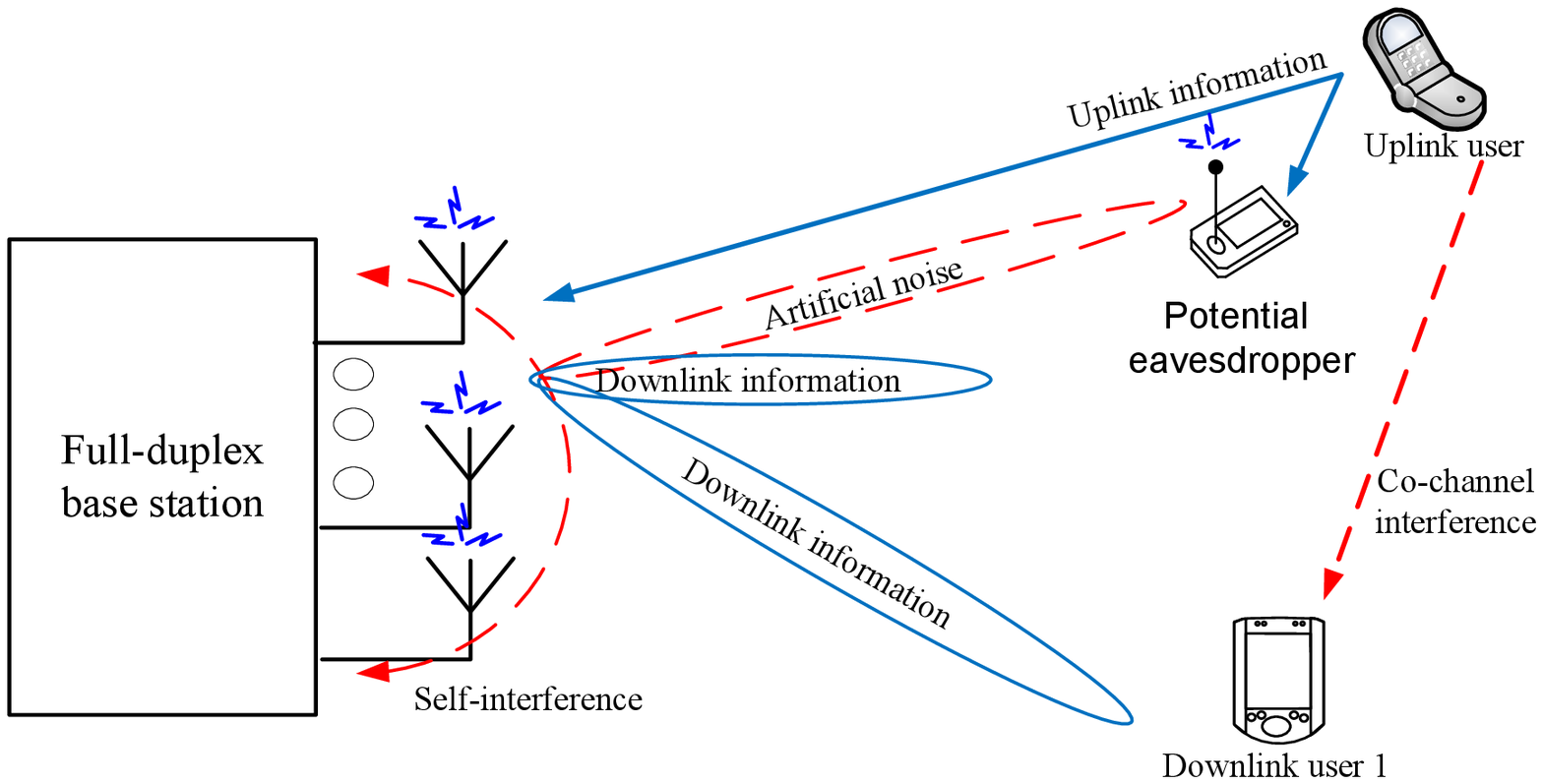}
\caption{Multiuser  system model  with an FD radio BS, $K=1$ HD downlink users, $J=1$ HD uplink users, and $M=1$ HD idle  receivers (potential eavesdroppers). The BS is equipped with $N$ antennas for facilitating secure  simultaneous uplink and downlink communication.}
\label{fig:system_model}
\end{figure}

\subsection{Notation}
We use boldface capital and lower case letters to denote matrices and vectors, respectively. $\mathbf{A}^H$, $\Tr(\mathbf{A})$, and $\Rank(\mathbf{A})$ represent the  Hermitian transpose, trace, and rank of  matrix $\mathbf{A}$, respectively; $\mathbf{A}^{-1}$ and $\mathbf{A}^{\dagger}$ represent the inverse and Moore-Penrose pseudoinverse of matrix $\mathbf{A}$, respectively; $\mathbf{A}\succeq \mathbf{0}$ indicates that $\mathbf{A}$ is a positive semidefinite matrix; $\mathbf{I}_N$ is the $N\times N$ identity matrix; $\mathbb{C}^{N\times M}$ denotes the set of all $N\times M$ matrices with complex entries; $\mathbb{H}^N$ denotes the set of all $N\times N$ Hermitian matrices; $\abs{\cdot}$ and $\norm{\cdot}$ denote the absolute value of a complex scalar and the Euclidean vector norm, respectively; ${\cal E}\{\cdot\}$ denotes statistical expectation;  $[x]^+=\max\{x,0\}$; the circularly symmetric complex Gaussian distribution with mean vector $\boldsymbol{\mu}$ and covariance matrix $\boldsymbol{\Sigma}$ is denoted by ${\cal CN}(\boldsymbol{\mu},\boldsymbol{\Sigma})$; and $\sim$ stands for ``distributed as".

\subsection{Multiuser System Model}
We consider a multiuser communication system which  consists of an FD radio BS, $K$ DL users, $J$ UL users, and $M$ idle users,  cf. Figure \ref{fig:system_model}.
 The FD radio BS is equipped with $N > 1$ antennas for simultaneous DL transmission and UL reception in the same frequency band via a circulator \cite{CN:FDRad}. The $K+J+M$ users are single-antenna HD mobile communication devices to ensure low hardware complexity. The DL and the UL users are scheduled for simultaneous DL and UL transmission while the $M$ idle users
are the receivers that are not scheduled in the current time slot. The signals intended for the DL users and the FD BS are overheard by the idle users. If the idle users are malicious, they may eavesdrop the emitted information
signals. Hence, the idle users are treated as
potential eavesdroppers in this paper which is taken into account for providing secure communication. Besides, we assume that the global channel state information (CSI) of all users is perfectly known at the BS for resource allocation\footnote{This work can be extended to the case of imperfect CSI knowledge of idle users (potential eavesdroppers) by following a similar approach as in \cite{JR:DAS_SWIPT,JR:MOOP}. }.
The number of antennas at the FD radio BS is assumed to be larger than the number of UL users and the number of idle users (potential eavesdroppers), respectively, i.e., $N>J$ and $N>M$, to facilitate UL signal detection and to guarantee communication security.

\subsection{Channel Model}
We consider a frequency flat fading channel.  In each scheduling time slot, the FD radio BS transmits $K$ independent signal streams simultaneously at the same frequency to the $K$ DL users. In particular, the information signal to DL user $k\in\{1,\ldots,K\}$ can be expressed as
\begin{eqnarray}
\mathbf{x}_k=\mathbf{w}_k d_k^{\mathrm{DL}},
\end{eqnarray}
where $d_k^{\mathrm{DL}}\in\mathbb{C}$ and $\mathbf{w}_k\in\mathbb{C}^{N\times1}$ are the information bearing signal for DL user $k$ and the corresponding DL beamforming vector, respectively.

 To provide secure communication in both DL and UL,  artificial noise is transmitted by the  FD radio BS and used to  interfere the reception of the
 potential eavesdroppers. Hence, the  DL transmit signal vector $\mathbf{x}$ at the  FD radio BS is given by
\begin{eqnarray}
\mathbf{x}=\underbrace{ \sum_{k=1}^K\mathbf{x}_k}_{\mbox{desired DL information signal}}+\underbrace{\mathbf{v}}_{\mbox{artificial noise}},
\end{eqnarray}
where $\mathbf{v}\in\mathbb{C}^{N\times 1}$ is the artificial noise vector generated by the  FD radio BS with distribution $\mathbf{v}\sim {\cal CN}(\mathbf{0}, \mathbf{V})$.

The received signals at DL user $k\in\{1,\ldots,K\}$, the FD radio BS, and idle user $m\in\{1,\ldots,M\}$ are given by
\begin{eqnarray}
\label{eqn:dl_user_rec_signal}
y_{k}^{\mathrm{DL}}\hspace*{-2mm}&=&\hspace*{-2mm}\mathbf{h}_k^H\mathbf{x}_k\hspace*{3mm}+\hspace*{-0.5mm}\underbrace{\sum_{i\neq k}^K\mathbf{h}_k^H\mathbf{x}_i}_{\mbox{multiuser interference}}+\underbrace{\mathbf{h}_k^H\mathbf{v}}_{\mbox{artificial noise}} \notag\\
& +& \hspace*{-0.5mm}\underbrace{\sum_{j=1}^J \sqrt{P_j}f_{j,k}d_j^{\mathrm{UL}}}_{\mbox{co-channel interference}}\hspace*{-0.5mm} +\hspace*{3mm} n^{\mathrm{DL}}_{k}\hspace*{-0.5mm},\,\, \\
\label{eqn:ul_rec_signal}\mathbf{y}^{\mathrm{UL}}\hspace*{-2mm}&=&\hspace*{-2mm}\sum_{j=1}^J \sqrt{P_j}\mathbf{g}_j d_j^{\mathrm{UL}}\hspace*{3mm}+\hspace*{-0.5mm} \underbrace{\mathbf{H}_{\mathrm{SI}}\sum_{k=1}^K{\mathbf{x}_k}}_{\mbox{self-interference}}\notag\\
&+&\underbrace{\mathbf{H}_{\mathrm{SI}}\mathbf{v}}_{\mbox{artificial noise}}
\hspace*{-0.5mm}+\hspace*{3mm}\mathbf{z},\,\quad \text{and},\\ \notag
y_{m}^{\mathrm{Eve}}\hspace*{-2mm}&=&\hspace*{-2mm}\sum_{k=1 }^K\mathbf{l}_m^H\mathbf{x}_k+\sum_{j=1}^J \sqrt{P_j}t_{j,m}d_j^{\mathrm{UL}}+\underbrace{\mathbf{l}_m^H\mathbf{v}}_{\mbox{artificial noise}} \\
\hspace*{-2mm}&+&\hspace*{-2mm}n^{\mathrm{Eve}}_{m},
\end{eqnarray}
respectively.  The channel between the FD radio BS and DL user $k$ is denoted by $\mathbf{h}_k\in\mathbb{C}^{N\times1}$ and $f_{j,k}\in\mathbb{C}$ represents the channel between UL user $j$ and DL user $k$. Variables $d_j^{\mathrm{UL}}$ and $P_j$ are the transmit data and the transmit power sent by UL user $j$ to the FD radio BS, respectively.  Without loss of generality, we assume ${\cal E}\{\abs{d_j^{\mathrm{UL}}}^2\}={\cal E}\{\abs{d_k^{\mathrm{DL}}}^2\}=1,\forall k\in\{1,\ldots,K\},j\in\{1,\ldots,J\}$. Vector $\mathbf{g}_j\in\mathbb{C}^{N\times1}$  denotes the channel between UL user $j$ and the FD radio BS.
Matrix $\mathbf{H}_{\mathrm{SI}}\in{\mathbb{C}^{N\times N}}$ denotes the self-interference (SI) channel which impairs the UL signal detection at the FD radio BS due to the concurrent DL transmission. Vector $\mathbf{l}_m\in\mathbb{C}^{N\times1}$   denotes the channel between the  FD radio BS and idle user $m$.  $t_{j,m}\in\mathbb{C}$ represents the channel between UL $j$ and idle user $m$. We note that variables $\mathbf{h}_k$, $f_{j,k}$, $\mathbf{g}_j$, $t_{j,m}$, and $\mathbf{H}_{\mathrm{SI}}$ capture the joint effects of path loss and small scale fading.  $\mathbf{z}\sim{\cal CN}(0,\sigma_{\mathrm{z}}^2\mathbf{I}_{N})$, $n^{\mathrm{DL}}_{k}\sim{\cal CN}(0,\sigma_{\mathrm{n}_k}^2)$, and $n^{\mathrm{Eve}}_{m}\sim{\cal CN}(0,\sigma_{\mathrm{Eve}_m}^2)$ represent the additive white Gaussian noise (AWGN) at the FD radio BS,  DL user $k$, and idle user $m$, respectively.

\begin{Remark}
We note that artificial noise is not generated by the single-antenna UL users due to their lack of spatial degrees of freedom. However, secure UL communication is facilitated by the artificial noise generated by the FD radio BS in the DL, as will be illustrated in the next section.
\end{Remark}

\section{Problem Formulation}\label{sect:forumlation}
In this section, we first introduce the QoS metrics for the considered FD radio communication system. Then, we formulate joint DL and UL power  allocation and beamforming design as a non-convex optimization problem.

\subsection{Achievable Throughput and Secrecy Rate}
The achievable throughput (bit/s/Hz) between the  FD radio BS and DL user $k\in\{1,\ldots,K\}$
 is given by
\begin{eqnarray}
C_k^{\mathrm{DL}}&=&\log_2(1+\Gamma^{\mathrm{DL}}_{k}),\\
\Gamma^{\mathrm{DL}}_{k}&=&\frac{\abs{\mathbf{h}_k^H\mathbf{w}_k}^2}{ S^{\mathrm{DL}}_k+\abs{\mathbf{h}_k^H\mathbf{v}}^2+ \sigma_{\mathrm{n}_k}^2},\,\,\\
 S^{\mathrm{DL}}_k&=&\overset{K}{\underset{m \neq k}{\sum}}\abs{\mathbf{h}_k^H\mathbf{w}_m}^2 + \overset{J}{\underset{j=1}{\sum}}P_j\abs{f_{j,k}}^2,
\end{eqnarray}
where $\Gamma^{\mathrm{DL}}_{k}$ is the receive SINR at DL user $k$.

Furthermore, we assume that the FD radio BS employs a linear receiver for decoding of the
received UL information for computational simplicity. Then, the achievable throughput between the  FD radio BS
and UL user $j\in\{1,\ldots,J\}$ is given by
\begin{eqnarray}
C_j^{\mathrm{UL}}&=&\log_2(1+\Gamma^{\mathrm{UL}}_{j}),\\
\Gamma^{\mathrm{UL}}_{j}&=&\frac{P_j\abs{\mathbf{g}_j^H\mathbf{r}_j}^2}{S^{\mathrm{UL}}_j+
\abs{\mathbf{r}_j^H\mathbf{H}_{\mathrm{SI}}\mathbf{v}}^2 +\sigma_{\mathrm{z}}^2\norm{\mathbf{r}_j}^2},\,\,\\
 S^{\mathrm{UL}}_j&=&\overset{J}{\underset{i \neq j}{\sum}}P_j\abs{\mathbf{g}_i^H\mathbf{r}_j}^2+\overset{K}{\underset{k=1}{\sum}}
\abs{\mathbf{r}_j^H\mathbf{H}_{\mathrm{SI}}\mathbf{w}_k}^2,
\end{eqnarray}
where $\Gamma^{\mathrm{UL}}_{j}$ is the receive SINR of UL user $j$ at the FD radio BS. Variable $\mathbf{r}_j\in\mathbb{C}^{N\times1}$ is the receive beamforming vector for decoding the information received from UL user $j$. In this paper, we assume that zero-forcing receive beamforming (ZF-BF) is adopted at the FD radio BS. We note that the performance of ZF-BF closely approaches that of minimum mean square error beamforming  when the noise term is not dominating \cite{book:david_wirelss_com} or the number of antennas at the FD radio BS is sufficiently large \cite{JR:FD_large_antennas}. Besides, ZF-BF facilitates a computational efficient resource allocation algorithm design.
Hence, the receive beamformer adopted at the FD radio BS for decoding the information transmitted by UL user $j$ is chosen as
 \begin{eqnarray}
 \mathbf{r}_j=(\mathbf{u}_j\mathbf{Q}^{\dagger})^H ,
  \end{eqnarray}
 where $\mathbf{u}_j=\Big[\underbrace{0,\ldots,0}_{(j-1)},1,\underbrace{0,\ldots,0}_{(J-j)}\Big]$, $\mathbf{Q}^{\dagger}=(\mathbf{Q}^H \mathbf{Q})^{-1}\mathbf{Q}^H$, and $\mathbf{Q}=[\mathbf{g}_1,\ldots,\mathbf{g}_J]$.

On the other hand,  we consider the worst-case scenario for providing secure communication in the DL. Specifically, in the worst case, idle user $m\in\{1,\ldots,M\}$ is able to
remove all DL multiuser interference and UL co-channel interference via successive interference cancellation  before attempting to decode the information of DL user $k$. Thus, the achievable throughput between the  FD radio BS and idle user (potential eavesdropper) $m$ for the message of DL user $k$  is given  by
\begin{eqnarray}\label{eqn:cap-eavesdropper}
C_{\mathrm{Eve}_m}^{\mathrm{DL}_k}\hspace*{-2mm}&=&\hspace*{-2mm}\log_2\Big(1\hspace*{-0.5mm}+\hspace*{-0.5mm}\Gamma_{\mathrm{Eve}_m}^{\mathrm{DL}_k}\Big)\,\,\,\,
\mbox{and}\,\,\\ \label{eqn:SINR_up_idle}
\Gamma_{\mathrm{Eve}_m}^{\mathrm{DL}_k}\hspace*{-2mm}&=&\hspace*{-2mm}\frac{\abs{\mathbf{l}_m^H\mathbf{w}_k}^2 }{\underset{j\ne k}{\sum}\abs{\mathbf{l}_m^H\mathbf{w}_j}^2 \hspace*{-0.5mm}+\hspace*{-0.5mm}\Tr(\mathbf{V}\mathbf{l}_m\mathbf{l}_m^H)\hspace*{-0.5mm}+\hspace*{-0.5mm}\overset{J}{\underset{j=1}{\sum}} P_j\abs{t_{j,m}}^2\hspace*{-0.5mm}+\hspace*{-0.5mm}\sigma_{\mathrm{Eve}_m}^2}\notag\\
&  \stackrel{(a)}{\le}& \frac{\abs{\mathbf{l}_m^H\mathbf{w}_k}^2 }{\Tr(\mathbf{V}\mathbf{l}_m\mathbf{l}_m^H)\hspace*{-0.5mm}+\hspace*{-0.5mm}\sigma_{\mathrm{Eve}_m}^2 }   ,
   \end{eqnarray}
where $\Gamma_{\mathrm{Eve}_m}^{\mathrm{DL}_k}$ is  the received SINR at idle user (potential eavesdropper) $m$, and $(a)$ reflects the aforementioned worst-case assumption which results in an upper bound on the received DL SINR at idle user (potential eavesdropper) $m$.

Besides, we also consider the worst-case scenario in guaranteeing communication secrecy for the UL users. Hence, we assume that the potential eavesdropper first removes  all UL multiuser interference and DL co-channel interference via successive interference cancellation  before attempting to decode the information of UL user $j$. Thus, the achievable throughput between  UL user $j$ and idle user (potential eavesdropper) $m$   is given  by
\begin{eqnarray}\label{eqn:cap-eavesdropper}
C_{\mathrm{Eve}_m}^{\mathrm{UL}_j}\hspace*{-2mm}&=&\hspace*{-2mm}\log_2\Big(1\hspace*{-0.5mm}+\hspace*{-0.5mm}\Gamma_{\mathrm{Eve}_m}^{\mathrm{UL}_j}\Big)\,\,\,\,
\mbox{and}\,\,\\ \label{eqn:SINR_up_idle}
\Gamma_{\mathrm{Eve}_m}^{\mathrm{UL}_j}\hspace*{-2mm}&=&\hspace*{-2mm}\frac{P_j\abs{t_{j,m}}^2 }{\overset{K}{\underset{m \neq k}{\sum}}\abs{\mathbf{l}_m^H\mathbf{w}_k}^2
\hspace*{-0.5mm}+\hspace*{-0.5mm}{\underset{i\ne j}{\sum}} P_i \abs{t_{i,m}}^2 \hspace*{-0.5mm}+\hspace*{-0.5mm}\Tr(\mathbf{V}\mathbf{l}_m\mathbf{l}_m^H)\hspace*{-0.5mm}+
\hspace*{-0.5mm}\sigma_{\mathrm{Eve}_m}^2}\notag\\
&  \stackrel{(b)}{\le}& \frac{P_j\abs{t_{j,m}}^2 }{\Tr(\mathbf{V}\mathbf{l}_m\mathbf{l}_m^H)\hspace*{-0.5mm}+\hspace*{-0.5mm}\sigma_{\mathrm{Eve}_m}^2 },
   \end{eqnarray}
where $\Gamma_{\mathrm{Eve}_m}^{\mathrm{UL}_j}$ is  the received SINR at idle user (potential eavesdropper) $m$, and $(b)$ reflects the aforementioned worst-case assumption which results in an upper bound on the received UL SINR at idle user (potential eavesdropper) $m$.

Thus, the achievable secrecy rates of
DL user $k$ and UL user $j$ are given by
\begin{eqnarray}\label{eqn:secure_cap}
C_{\mathrm{sec}}^{\mathrm{DL}_k}&=&\Big[C_k^{\mathrm{DL}}-\underset{\forall m}{\max}\,\{ C_{\mathrm{Eve}_m}^{\mathrm{DL}_k}\}\Big]^+ \mbox{and}\\
C_{\mathrm{sec}}^{\mathrm{UL}_j}&=&\Big[C_j^{\mathrm{UL}}-\underset{\forall j}{\max}\,\{C_{\mathrm{Eve}_m}^{\mathrm{UL}_j}\}\Big]^+,
\end{eqnarray}
respectively.
\begin{Remark}
We note that a FD radio BS can provide communication security also to UL users. Specifically, the FD radio BS not only decodes the UL information, but also transmits artificial noise in the DL concurrently to interfere the potential eavesdroppers, cf. \eqref{eqn:SINR_up_idle}, which is not possible for traditional UL communication served by a HD radio BS.
\end{Remark}

\subsection{Optimization Problem Formulation}
\label{sect:cross-Layer_formulation}
The system objective is to minimize the weighted sum of the DL and UL transmit powers while providing
QoS for reliable and secure communication to both DL and UL users simultaneously.
The optimal power allocation and beamformer design are obtained by solving the following optimization problem:
\begin{eqnarray}
\label{eqn:prob1}
&&\hspace*{-10mm} \underset{\mathbf{w}_k,\mathbf{v},P_j}{\mino}\,\, \,\, \alpha\Big(\sum_{k=1}^{K}\norm{\mathbf{w}_k}^2+\norm{\mathbf{v}}^2\Big) +\beta\sum_{j=1}^J P_j \notag \vspace*{-0.5cm}\\
\notag\mathrm{s.t.}
&&\hspace*{-5mm}\mbox{C1: }\frac{\abs{\mathbf{h}_k^H\mathbf{w}_k}^2}{S^{\mathrm{DL}}_k+\abs{\mathbf{h}_k^H\mathbf{v}}^2+ \sigma_{\mathrm{n}_k}^2 } \geq \Gamma^{\mathrm{DL}}_{\mathrm{req}_k},\,\, \forall k, \notag\\
&&\hspace*{-5mm}\mbox{C2: }\frac{P_j\abs{\mathbf{g}_j^H\mathbf{r}_j}^2}{S^{\mathrm{UL}}_j +\sigma_{\mathrm{z}}^2\norm{\mathbf{r}_j}^2} \geq \Gamma^{\mathrm{UL}}_{\mathrm{req}_j},\,\, \forall j, \notag\\
&&\hspace*{-5mm}\mbox{C3: }
\frac{\abs{\mathbf{l}_m^H\mathbf{w}_k}^2 }{\Tr(\mathbf{V}\mathbf{l}_m\mathbf{l}_m^H)\hspace*{-0.5mm}+\hspace*{-0.5mm}\sigma_{\mathrm{Eve}_m}^2 } \leq \Gamma_{\mathrm{tol}_m}^{\mathrm{DL}_k},\forall m,k,\notag\\
&&\hspace*{-5mm}\mbox{C4: }
\frac{P_j\abs{t_{j,m}}^2 }{\Tr(\mathbf{V}\mathbf{l}_m\mathbf{l}_m^H)\hspace*{-0.5mm}+\hspace*{-0.5mm}\sigma_{\mathrm{Eve}_m}^2 }\leq \Gamma_{\mathrm{tol}_m}^{\mathrm{UL}_j},\forall m,j,\notag\\
&&\hspace*{-5mm}\mbox{C5: } P_j \geq 0,\,\, \forall j.
\end{eqnarray}
Constant variables $\alpha,\beta\geq 0$ in the objective function reflect the preference of the system operator for reducing the
DL transmit power and UL transmit power, respectively. Besides, $\Gamma^{\mathrm{DL}}_{\mathrm{req}_k} > 0$ and $\Gamma^{\mathrm{UL}}_{\mathrm{req}_j} > 0$ are the minimum required SINRs for DL user $k\in\{1,\ldots,K\}$ and UL user $j\in\{1,\ldots,J\}$, respectively.  Constraints C3 and C4 are imposed such that  the maximum received SINR at idle user (potential eavesdropper)
$m$ is less than the maximum tolerable received SINRs $\Gamma_{\mathrm{tol}_m}^{\mathrm{DL}_k}$ and $\Gamma_{\mathrm{tol}_m}^{\mathrm{UL}_j}$, when idle user $m$ attempts to decode the information of DL user $k$ and UL  user $j$, respectively. In practice, the service provider selects $\Gamma^{\mathrm{DL}}_{\mathrm{req}_k}\gg  \Gamma_{\mathrm{tol}_m}^{\mathrm{DL}_k}$ and $\Gamma^{\mathrm{UL}}_{\mathrm{req}_j}\gg \Gamma_{\mathrm{tol}_m}^{\mathrm{UL}_j},\forall m\in\{1,\ldots,M\}, j, k,$  for providing communication security in the DL and UL, respectively.  In other words, the FD BS is able to guarantee minimum secrecy rates of $C_{\mathrm{sec}}^{\mathrm{DL}_k}\ge \log_2(1+\Gamma^{\mathrm{DL}}_{\mathrm{req}_k})-\log_2(1+\Gamma_{\mathrm{tol}_m}^{\mathrm{DL}_k})$ and $C_{\mathrm{sec}}^{\mathrm{UL}_j}\ge \log_2(1+\Gamma^{\mathrm{UL}}_{\mathrm{req}_j})-\log_2(1+\Gamma_{\mathrm{tol}_m}^{\mathrm{UL}_j})$  for DL and UL, respectively.  Constraint C5 in (\ref{eqn:prob1}) is the non-negative power constraint for UL user $j$.

\section{Solution of the Optimization Problem} \label{sect:solution}
The optimization problem in (\ref{eqn:prob1}) is a non-convex problem due to the non-convexity of constraints C1 and C2.  In general, there is no systematic approach for handling non-convex optimization problems. In some cases, an exhaustive search over the feasible solution set is needed to obtain the global optimal solution which often entails an exponential computational complexity. In order to solve the problem efficiently, we recast  \eqref{eqn:prob1} as a convex optimization problem via SDP relaxation and verify the optimality of the proposed relaxation.
For facilitating  SDP relaxation, we first define the following auxiliary variable matrices:
\begin{eqnarray}\label{eqn:change_of_variables}
\mathbf{W}_k&=&\mathbf{w}_k\mathbf{w}_k^H,\,\mathbf{H}_k=\mathbf{h}_k\mathbf{h}_k^H,\,\mathbf{G}_j=\mathbf{g}_j\mathbf{g}_j^H,\notag\\
\mathbf{R}_j&=&\mathbf{r}_j\mathbf{r}_j^H, \mathbf{L}_m=\mathbf{l}_m\mathbf{l}_m^H,
\end{eqnarray}
and rewrite \eqref{eqn:prob1} in the following equivalent form:
\begin{eqnarray}
\label{eqn:sdp1}&&\hspace*{-10mm}\underset{{\mathbf{W}_k},\,\mathbf{V}\in \mathbb{H}^{N},P_j
}{\mino}\,\, \alpha\Big(\sum_{k=1}^{K}\Tr(\mathbf{W}_k)+\Tr(\mathbf{V})\Big) +\beta\sum_{j=1}^J P_j\nonumber\\
\notag \mathrm{s.t.} &&\hspace*{-5mm}{\text{C1}}\mbox{: } \notag\frac{\Tr(\mathbf{H}_k\mathbf{W}_k)}{\Gamma^{\mathrm{DL}}_{\mathrm{req}_k}} \ge {I_k^{\mathrm{DL}}+\Tr(\mathbf{H}_k\mathbf{V})+\sigma_{{\mathrm{n}}_k}^2},\,\, \forall k, \\
&&\hspace*{-5mm}{\text{C2}}\mbox{: }\notag\frac{P_j\Tr(\mathbf{R}_j\mathbf{G}_j)}{\Gamma^{\mathrm{UL}}_{\mathrm{req}_j}}
 \ge {I_j^\mathrm{UL}+\sigma_{\mathrm{z}}^2\Tr(\mathbf{R}_j)},\,\, \forall j,\end{eqnarray}\begin{eqnarray}
 &&\hspace*{-5mm}\mbox{C3: }
\frac{\Tr(\mathbf{L}_m\mathbf{W}_k) }{\Gamma_{\mathrm{tol}_m}^{\mathrm{DL}_k}} \leq  \Tr(\mathbf{L}_m\mathbf{V})\hspace*{-0.5mm}+\hspace*{-0.5mm}\sigma_{\mathrm{Eve}_m}^2 ,\forall m,k,\notag\\
&&\hspace*{-5mm}\mbox{C4: }
\frac{P_j\abs{t_{j,m}}^2 }{\Gamma_{\mathrm{tol}_m}^{\mathrm{UL}_j}}\leq \Tr(\mathbf{L}_m\mathbf{V})\hspace*{-0.5mm}+\hspace*{-0.5mm}\sigma_{\mathrm{Eve}_m}^2  ,\forall m,j,\notag\\
&&\hspace*{-10mm}{\text{C5}},\, {\text{C6}}\mbox{: }\,\, \mathbf{W}_k\succeq \mathbf{0},\,\, \forall k,\,{\text{C7}}\mbox{: }\,\, \Rank(\mathbf{W}_k) \le 1,\,\, \forall k,
\end{eqnarray}
where $\mathbf{W}_k\succeq \mathbf{0}$, ${\mathbf{W}_k}\in \mathbb{H}^{N}$, and $\Rank(\mathbf{W}_k) \le 1$ in (\ref{eqn:sdp1}) are imposed to guarantee that $\mathbf{W}_k=\mathbf{w}_k\mathbf{w}_k^H$ holds after optimization, and
\begin{eqnarray}
I_k^\mathrm{DL}\hspace*{-2.5mm}&=&\hspace*{-2.5mm} \overset{K}{\underset{i \neq k}{\sum}} \Tr(\mathbf{H}_k
\mathbf{W}_i)+\overset{J}{\underset{j=1}{\sum}} P_j\abs{f_{j,k}}^2,\,\, \text{and}\\
I_j^\mathrm{UL}\hspace*{-2.5mm}&=&\hspace*{-2.5mm}\overset{J}{\underset{i \neq j}{\sum}} P_r\Tr(\mathbf{R}_j
\mathbf{G}_i)+ \Tr\Big(\big(\mathbf{V}+\overset{K}{\underset{k=1}{\sum}}\mathbf{W}_k\big)\mathbf{H}_{\mathrm{SI}}^H\mathbf{R}_j\mathbf{H}_{\mathrm{SI}}\Big)\notag.
\end{eqnarray}
Transformed optimization problem \eqref{eqn:sdp1} is a non-convex problem due to the combinatorial rank-one constraint ${\text{C7}}$. In order to obtain a tractable solution, we adopt  constraint relaxation by removing ${\text{C7}}$ from the problem formulation which yields:
  \begin{eqnarray}
\label{eqn:sdp_relaxed}&&\hspace*{-10mm}\underset{{\mathbf{W}_k},\,\mathbf{V}\in \mathbb{H}^{N},P_j
}{\mino}\,\, \alpha\Big(\sum_{k=1}^{K}\Tr(\mathbf{W}_k)+\Tr(\mathbf{V})\Big) +\beta\sum_{j=1}^J P_j\notag\\
\mathrm{s.t.} &&\hspace*{15mm}{\text{C1 -- C6}}.
\end{eqnarray}
The SDP relaxed convex problem in (\ref{eqn:sdp_relaxed}) can be solved efficiently by standard interior point methods. Next, we reveal a sufficient condition for obtaining a rank-one solution $\mathbf{W}_k$ for \eqref{eqn:sdp_relaxed} in the following theorem.

\begin{Thm}\label{thm:rankone_condition} If the channel vectors of the DL users, $\mathbf{h}_k,k\in\{1,\ldots,K\},$ the UL users, $\mathbf{g}_j,j\in\{1,\ldots,J\},$  and the potential eavesdroppers, $\mathbf{l}_m,m\in\{1,\ldots,M\},$ as well as the SI interference channel matrix $\mathbf{H}_{\mathrm{SI}}$ can be modeled as statistically independent random variables, the solution of (\ref{eqn:sdp_relaxed}) is rank-one, i.e.,  $\Rank(\mathbf{W}_k)=1$ for $\mathbf{W}_k\ne\mathbf{0}, \, \forall k$, with probability one.
\end{Thm}

\emph{\quad Proof: }Please refer to the Appendix. \qed

In other words, the optimal beamformer $\mathbf{w}^*_k$ in \eqref{eqn:prob1} can be obtained  by performing eigenvalue decomposition of
$\mathbf{W}_k$, if the channels satisfy the condition stated in Theorem $1$.
\section{Results}
In this section, we verify the performance of the proposed optimal power and beamforming resource allocation scheme through simulations.  The relevant simulation parameters are summarized in Table \ref{tab:parameters}. There are $K=6$ DL users, $J=3$ UL users, and $M=5$ idle users (potential eavesdroppers) in the cell. All users are  randomly and uniformly distributed between the reference distance and the  maximum service distance of $500$ meters. The FD radio BS is located at the center of the system which is equipped with $N$ antennas. The small scale fading of the DL channels, UL channels, and inter-user channels is modeled as independent and identically distributed Rayleigh fading. The multipath fading coefficients of the SI channel are generated as independent and identically distributed Rician random variables with Rician factor $6$ dB. Besides, we assume $\alpha=\beta=1$ to study the system performance. Also,  the UL users require a fixed minimum SINR of $10$ dB, i.e., $\Gamma^{\mathrm{UL}}_{\mathrm{req}_j}=10$ dB, $\forall j\in\{1,\ldots,J\}$, while  the DL users require identical minimum SINRs, i.e., $\Gamma^{\mathrm{DL}}_{\mathrm{req}_k}=\Gamma^{\mathrm{DL}}_{\mathrm{req}},\forall k\in\{1,\ldots,K\}$.

\begin{table}[t]\caption{System parameters.}\label{tab:parameters} 
\newcommand{\tabincell}[2]{\begin{tabular}{@{}#1@{}}#2\end{tabular}}
\centering
\begin{tabular}{|l|l|}\hline
\hspace*{-1mm}Carrier center frequency & $1.9$ GHz  \\
\hline
\hspace*{-1mm}System bandwidth & $200$ KHz  \\
\hline
\hspace*{-1mm}Path loss exponent &  $3.6$  \\
\hline
\hspace*{-1mm}Reference distance &  $30$ m  \\
\hline
\hspace*{-1mm}Maximum tolerable receive SINR &  $\Gamma_{\mathrm{tol}}=\Gamma_{\mathrm{tol}_m}^{\mathrm{DL}_k}=\Gamma_{\mathrm{tol}_m}^{\mathrm{UL}_j}=-10$ dB  \\
\hline
\hspace*{-1mm}SI cancellation    &  $-110$ dB \cite{CN:FDRad}   \\
\hline
\hspace*{-1mm}Thermal noise power &  $-121$ dBm   \\
\hline
\hspace*{-1mm}DL and idle user noise figure &$9$ dB   \\
\hline
\hspace*{-1mm}BS noise figure &  $2$ dB  \\
\hline
\hspace*{-1mm}BS antenna gain &  $18$ dBi (decibel isotropic)  \\
\hline
\end{tabular}
\end{table}

\subsection{Average Total Transmit Power}
\begin{figure}[t]
 \centering \vspace*{-3mm}
\includegraphics[width=3.5in]{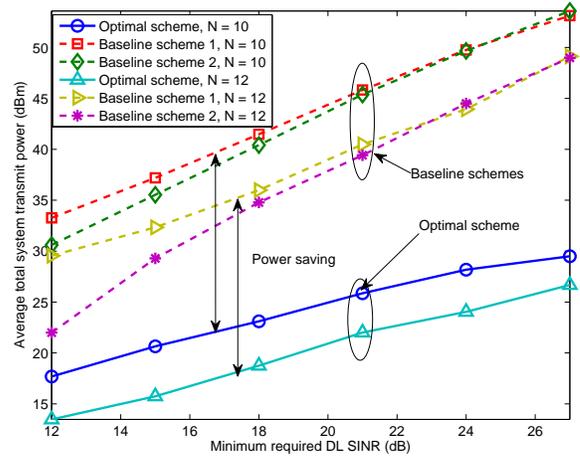}
\caption{Average total system transmit power (dBm) versus the minimum required DL SINR, $\Gamma^{\mathrm{DL}}_{\mathrm{req}}$ (dB), for different resource allocation schemes. The double-sided arrows indicate the power saving enabled by the propose optimal resource allocation scheme.} \label{fig:pt_SINR}
\end{figure}

In Figure \ref{fig:pt_SINR}, we study the average total system transmit power versus the minimum required DL SINR, $\Gamma_{\mathrm{req}}^{\mathrm{DL}}$, for   different numbers of antennas at the FD radio BS. It can be observed from Figure \ref{fig:pt_SINR} that the total  transmit power is a monotonically increasing function with respect to the minimum required DL SINR. In particular, as the minimum  DL SINR requirement becomes more stringent, the FD radio BS allocates more  power to the DL information signals. Meanwhile, more artificial noise power is also needed to neutralize the information leakage to the potential eavesdroppers. The higher DL transmit power leads in turn to a higher SI which impairs the UL transmission. Thus, the UL users also have to increase their transmit power in order to meet the minimum UL SINR requirements which results in  an increase of the total system transmit power. On the other hand, Figure \ref{fig:pt_SINR} reveals that the total average system transmit power
decreases with increasing number of FD radio BS antennas since extra
degrees of freedom can be exploited for DL resource allocation and UL signal detection,
when more antennas are available.

For comparison, we also consider two baseline resource allocation schemes.  For
baseline scheme $1$, we perform maximum ratio transmission for the artificial noise with respect to the virtual channel spanned by
the idle users, i.e., $\mathbf{L}=[\mathbf{l}_1\ldots,\mathbf{l}_M]$. Then, we minimize the total transmit power by optimizing $\mathbf{W}_k$, the power of the artificial noise, and $P_j$  subject to constraints C1-C5 as in \eqref{eqn:prob1} via SDP relaxation.  Baseline
scheme $2$ has the same structure as baseline scheme $1$ except
that the artificial noise is radiated isotropically. It can be
observed that the two baseline schemes requires a significantly higher total transmit power than the proposed optimal scheme. Indeed,
the proposed optimal scheme fully utilizes the CSI of
all communication links and optimizes the space spanned by
the artificial noise for providing secure and reliable communication. On the contrary,
the direction of the artificial noise signal is fixed in the two baseline schemes leading to a less effective
jamming of the potential eavesdroppers and more severe SI at the FD radio BS.

\begin{figure}[t]
 \centering \vspace*{-3mm}
\includegraphics[width=3.5in]{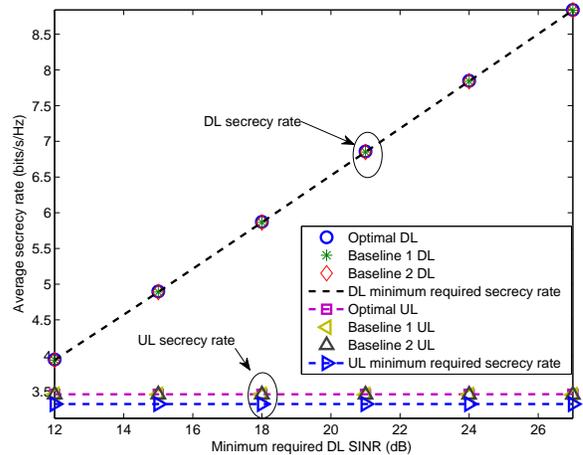}
\caption{Average secrecy rate (bit/s/Hz) versus the minimum required DL SINR, $\Gamma^{\mathrm{DL}}_{\mathrm{req}}$ (dB), for different resource allocation schemes.} \label{fig:cap_SNR}\vspace*{-3mm}
\end{figure}

\subsection{Average Secrecy Rate}
Figure \ref{fig:cap_SNR} illustrates the average  secrecy rate (bit/s/Hz) versus the minimum required DL SINR, $\Gamma^{\mathrm{DL}}_{\mathrm{req}}$, for different resource allocation schemes and $N=8$ FD radio BS antennas.  It can be seen that the average DL secrecy rate, i.e., $C_{\mathrm{sec}}^{\mathrm{DL}_k}\ge \log_2(1+\Gamma^{\mathrm{DL}}_{\mathrm{req}})-\log_2(1+\Gamma_{\mathrm{tol}})$, increases with  $\Gamma^{\mathrm{DL}}_{\mathrm{req}}$
since the maximum tolerable SINRs of the idle users  are constrained to be less than $\Gamma_{\mathrm{tol}}=-10$ dB.
On the other hand, although the increased minimum required DL SINR leads to a higher SI interference power impairing the UL signal detection,  the total average secrecy rate for the UL communication still satisfies the minimum requirement and remains  constant due to the proposed optimal resource allocation. Besides, all  considered schemes achieve the same secrecy rate. However, the proposed optimal scheme consumes much less power than the baseline schemes to achieve this secrecy rate, cf. Figure \ref{fig:pt_SINR}.

\begin{Remark}
The performance of DL  and UL communication with a HD radio BS is not shown in this paper. In fact, a HD radio BS cannot guarantee secure UL communication due to the lack of spatial degrees of freedom at each UL user for jamming with artificial noise. Hence,   the optimization problem for the HD radio BS corresponding to \eqref{eqn:prob1} is always infeasible.
\end{Remark}

\section{Conclusions}\label{sect:conclusion}
In this paper, we formulated the joint power allocation and beamforming design for simultaneous DL and UL wireless communication via an  FD radio BS as a non-convex optimization problem. The problem formulation took into account
communication security for both DL and UL transmission and artificial
noise injection at the FD radio BS. A power efficient SDP based resource allocation
scheme was proposed to obtain the optimal solution for
minimization of the weighted sum of the DL and UL transmit powers.  Simulation results  unveiled the power savings enabled by optimization of the artificial noise covariance matrix. Besides, we showed that unlike a HD BS,  an FD BS can  guarantee communication
security for DL and UL users simultaneously.

\section*{Appendix - Proof of Theorem 1}
It can be verified that the problem in (\ref{eqn:sdp_relaxed}) is jointly convex with respect to the optimization variables and satisfies the Slater's constraint qualification. As a result, the duality gap is zero and  solving the dual problem is equivalent to solving the primal problem \cite{book:convex}. Hence, we start the proof by writing the Lagrangian function of the primal problem in \eqref{eqn:sdp_relaxed} in terms of the beamforming matrix $\mathbf{W}_k$:
\begin{eqnarray}
{\cal L}\Big(\mathbf{\Theta},\mathbf{\Phi}\Big)\hspace*{-3mm}&=&\hspace*{-3mm}\sum_{k=1}^{K}\Tr(\mathbf{B}_k\mathbf{W}_k)-
\sum_{k=1}^{K}\Tr\Big(
\big(\mathbf{Y}_k+\frac{\delta_k\mathbf{H}_{k}}{\Gamma_{\mathrm{req}_k}^{\mathrm{DL}}}\big)\mathbf{W}_k\Big)\notag\\
\hspace*{-3mm}&+&\hspace*{-3mm}\Delta \\
\label{eqn:B_k}
\mbox{and}\quad\mathbf{B}_k\hspace*{-3mm}&=&\hspace*{-3mm}\alpha\mathbf{I}_{N}+\sum_{j\neq k}^{K}\delta_j\mathbf{H}_{j}+\sum_{j=1}^{J}\gamma_j\mathbf{H}_{\mathrm{SI}}^H\mathbf{R}_j\mathbf{H}_{\mathrm{SI}}\notag\\
\hspace*{-3mm}&+&\hspace*{-3mm}\sum_{m=1}^M \lambda_{m,k}\frac{\mathbf{L}_m}{\Gamma_{\mathrm{tol}_m}^{\mathrm{DL}_k}},
\end{eqnarray}
where $\mathbf{\Theta}\triangleq\{\mathbf{W}_k,\mathbf{V},P_j\}$ and $\mathbf{\Phi}\triangleq\{\delta_k,\gamma_j,\lambda_{m,k},\mathbf{Y}_k\}$ are the sets of primal and dual variables, respectively. $\delta_k,\gamma_j,\lambda_{m,k}\ge 0$ and  $\mathbf{Y}_k\succeq \zero$ are the dual variables with respect to constraints C1, C2, C3, and C6 in \eqref{eqn:sdp_relaxed}, respectively.
$\Delta$ denotes the collection of  variables that are independent of $\mathbf{W}_k$. For convenience,  the optimal primal and  dual variables of  (\ref{eqn:sdp_relaxed}) are denoted by the corresponding variables with an asterisk  superscript. We focus on the following Karush-Kuhn-Tucker (KKT) optimality conditions:
\begin{eqnarray}\vspace*{-2mm}
\hspace*{-3mm}\mathbf{Y}_k^*\hspace*{-3mm}&\succeq&\hspace*{-3mm}\mathbf{0},\,\,\delta_k^*\ge 0,\,\forall k, \label{eqn:dual_variables}\\
\hspace*{-3mm}\mathbf{Y}_k^*\mathbf{W}_k^*\hspace*{-3mm}&=&\hspace*{-3mm}\mathbf{0},\label{eqn:KKT-complementarity}\\
\hspace*{-3mm}\mathbf{Y}_k^*\hspace*{-3mm}&=&\hspace*{-3mm}\mathbf{B}_{k}^*-\frac{\delta_k^*\mathbf{H}_k}
{\Gamma_{\mathrm{req}_k}^{\mathrm{DL}}},
\label{eqn:lagrangian_gradient}
\end{eqnarray}
where $\mathbf{B}_{k}^*$ in (\ref{eqn:lagrangian_gradient}) is obtained by substituting the optimal dual variables $\boldsymbol \Phi^*$ into (\ref{eqn:B_k}). From (\ref{eqn:KKT-complementarity}), we know that the optimal beamforming matrix $\mathbf{W}^*_k$ is a rank-one matrix when  $\Rank(\mathbf{Y}^*_k)=N-1$. In particular, $\mathbf{W}^*_k$ is required to lie in the null space spanned by $\mathbf{Y}^*_k$ for $\mathbf{W}^*_k\ne\zero$. As a result, by revealing the structure of $\mathbf{Y}^*_k$, we can study the rank of beamforming matrix $\mathbf{W}^*_k$. In the following, we first show by contradiction that $\mathbf{B}_k^*$ is a positive definite matrix with probability one. To this end, we focus on the dual problem of (\ref{eqn:sdp_relaxed}). For a given set of optimal dual variables, $\mathbf{\Phi}^*=\{\delta^*,\lambda_{m,k}^*,\gamma_j^*,\mathbf{Y}^*_k\}$,  and a subset of optimal primal variables, $\{P^{*}_j,\mathbf{V}^*\}$,  the dual problem of (\ref{eqn:sdp_relaxed}) can be written as
\begin{eqnarray}\hspace*{-2mm}\label{eqn:dual2}
\,\,\underset{\mathbf{W}_k\in\mathbb{H}^{N}}{\mino} \,\, {\cal L}\Big(\hspace*{-0.5mm}\mathbf{\Theta},\mathbf{\Phi}^*\hspace*{-0.5mm}\Big).
\end{eqnarray}
Suppose $\mathbf{B}_k^*$ is negative semi-definite, i.e., $\mathbf{B}_k^*\preceq\zero$,  then we can construct a beamforming matrix $\mathbf{W}_k=s\mathbf{\tilde w}_k\mathbf{\tilde w}_k^H$ as one of the feasible solutions of (\ref{eqn:dual2}), where $s>0$ is a scaling parameter and $\mathbf{\tilde w}_k$ is the eigenvector corresponding to one of the non-positive eigenvalues of $\mathbf{B}_k^*$. We substitute $\mathbf{W}_k=s\mathbf{\tilde w}_k\mathbf{\tilde w}_k^H$ into (\ref{eqn:dual2}) which yields
\begin{eqnarray}\vspace*{-2mm}\notag
{\cal L}\Big(\mathbf{\Theta},\mathbf{\Phi}\Big)&=&\underbrace{\sum_{k=1}^{K}\Tr(s\mathbf{B}_k^*\mathbf{\tilde w}_k\mathbf{\tilde w}_k^H)}_{\le 0}\end{eqnarray}\begin{eqnarray}
&-&r\sum_{k=1}^{K}\Tr\Big(\mathbf{\tilde w}_k\mathbf{\tilde w}_k^H
\big(\mathbf{Y}_k^*+\frac{\delta_k^*\mathbf{H}_k}{\Gamma_{\mathrm{req}_k}^{\mathrm{DL}}}\big)\Big)+\Delta.
\end{eqnarray}
 Besides, by using a similar approach as in \cite{JR:DAS_SWIPT}, it can be shown that constraint C1 is satisfied with equality for the optimal solution and thus $\delta_k>0$. Furthermore,  since the channel vectors of the DL users, i.e.,   $\mathbf{H}_{k}$, $\forall k\in\{1,\ldots,K\}$, are assumed to be statistically independent of the other channels in the system, we obtain $-r\sum_{k=1}^{K}\Tr\Big(\mathbf{\tilde w}_k\mathbf{\tilde w}_k^H
\big(\mathbf{Y}_k^*+\frac{\delta_k^*\mathbf{H}_{k}}{\Gamma_{\mathrm{req}_k}^{\mathrm{DL}}}\big)\Big)\rightarrow -\infty$ for $r\rightarrow \infty$. Thus, the dual optimal value  becomes unbounded from below. Yet, the optimal value of the primal problem in (\ref{eqn:sdp_relaxed}) is non-negative which leads to a contradiction as strong duality would not hold. Therefore, for the optimal solution,  $\mathbf{B}_k^*$ has to be a positive definite and full rank matrix with probability one, i.e., $\Rank(\mathbf{B}_k^*)=N$.

Furthermore, we have the following implication:
\begin{subequations}
\begin{eqnarray}
\hspace*{-3mm}&&\Rank(\mathbf{Y}^*_k)+\Rank\big(\delta_k^*\frac{\mathbf{H}_{k}}{\Gamma_{\mathrm{req}_k}^{\mathrm{DL}}}\big)\\
\hspace*{-3mm}&\stackrel{(c)}{\ge}&\Rank\big(\mathbf{Y}^*_k+\delta_k^*\frac{\mathbf{H}_{k}}{\Gamma_{\mathrm{req}_k}^{\mathrm{DL}}}
\big)\\
&\stackrel{(d)}{=}&\Rank(\mathbf{B}_k^*)=N\Rightarrow
 \Rank(\mathbf{Y}^*_k)\ge N-1,
\end{eqnarray}
\end{subequations}
where $(c)$ and $(d)$ are due to a basic rank inequality and  (\ref{eqn:lagrangian_gradient}), respectively.
 Furthermore, $\mathbf{W}_k^*\ne\mathbf{0}$ is required to satisfy  C1 for $\Gamma_{\mathrm{req}_k}^{\mathrm{DL}}>0$. Thus, $\Rank(\mathbf{Y}^*_k)=N-1$ and $\Rank(\mathbf{W}^*_k)=1$ hold with probability one.  \qed


\end{document}